\documentclass[useAMS,usenatbib]{mn2e}
\usepackage{graphicx}
\usepackage{times}
\usepackage[dvips]{color}

\def\ebv{$E(B-V)$~}

\def\msun{M$_{\odot}$}
\def\gsim{\;\lower.6ex\hbox{$\sim$}\kern-7.75pt\raise.65ex\hbox{$>$}\;}
\def\lsim{\;\lower.6ex\hbox{$\sim$}\kern-7.75pt\raise.65ex\hbox{$<$}\;}

\title[Berkeley 32 and King 11]{The old open clusters 
Berkeley 32 and King 11\thanks{ 
Based on observations made with the Italian
Telescopio Nazionale Galileo (TNG) operated on the island of La Palma by
the Fundaci\'on Galileo Galilei of the INAF (Istituto Nazionale di
Astrofisica) at the Spanish Observatorio del Roque de los Muchachos of
the Instituto de Astrofisica de Canarias.}
}

\author[Tosi et al.]{
 Monica Tosi$^{1}$,\thanks{E-mail: 
  monica.tosi@oabo.inaf.it (MT), 
  angela.bragaglia@oabo.inaf.it (AB), 
  michele.cignoni@unibo.it (MC)} 
 Angela Bragaglia$^{1}$ and Michele Cignoni$^{1,2}$\\
 \\
 $^{1}$ INAF--Osservatorio Astronomico di Bologna, Via Ranzani 1, I-40127 Bologna
      (Italy) \\
 $^{2}$ Dipartimento di Astronomia, Universit\`a di Bologna, via Ranzani 1, 
  I-40127 Bologna (Italy)\\
     }

\date{}

\begin{document}
\maketitle

\begin{abstract}
We have obtained CCD $BVI$ imaging of the old open clusters Berkeley~32 and 
King 11.  Using the synthetic colour-magnitude diagram method with
three different sets of stellar evolution models of various metallicities, with and without overshooting, we have determined their age, 
distance, reddening, and indicative metallicity, as well as distance from the
Galactic centre and height from the Galactic plane.   
The best parameters derived for Berkeley~32 are: 
subsolar metallicity (Z=0.008 represents the
best choice, Z=0.006 or 0.01 
are more marginally acceptable), age = 5.0--5.5 Gyr (models with overshooting;
without overshooting the age is 4.2--4.4 Gyr with poorer agreement), 
$(m-M)_0=12.4-12.6$, $E(B-V)=0.12-0.18$ (with the lower value being more
probable because it corresponds to the best metallicity), 
$R_{GC}\sim 10.7-11$~kpc, and $|Z|\sim 231-254$~pc. 
The best parameters for King~11 are:  Z=0.01,
age=3.5--4.75 Gyr, $(m-M)_0=11.67-11.75$, $E(B-V)=1.03-1.06$, $R_{GC}\sim
9.2-10$~kpc, and $|Z|\sim 253-387$~pc.

\end{abstract}


\section{Introduction}

This paper is part of the BOCCE (Bologna Open Cluster Chemical Evolution) 
project, described in detail by \cite{bt06}. With this project, we intend to
derive homogeneous measures of age, distance, reddening and chemical
abundance for a large sample of open clusters (OCs), to study the
present day properties of the Galactic disc and their evolution with
time.

As part of this project, we present here a photometric study of  the two
old OCs King~11
($\alpha_{2000}=23^h47^m40^s$, 
$\delta_{2000}=+68^\circ38\arcmin30\arcsec$, $l=117.^\circ2$, $b=+6.^\circ5$)
and Berkeley~32 
($\alpha_{2000}=06^h58^m07^s$,
$\delta_{2000}=+06^\circ 25\arcmin43\arcsec$, $l=208^\circ$,  $b=+4.4^\circ$),
located in the second and third Galactic quadrants, respectively.

King~11 has been the subject of a few publications in the past.
\cite{kaluzny89} obtained a  rather shallow colour-magnitude diagram (CMD)
using the 0.9m KPNO telescope. He found it old (about the same age of M~67) 
and highly reddened, with a distance modulus ($m-M)_V \sim 15.3$,  
derived assuming  $M_V$(clump)=0.7 mag.
\cite{aparicio91} acquired deep $UBVR$ data at the 3.5m telescope in
Calar Alto on a small field of view (2.7$\times$4.3 arcmin$^2$); they derived 
a reddening \ebv=1, a distance modulus $(m-M)_0\simeq11.7$, a metallicity
about solar (with some uncertainty, because different methods produced
contrasting answers), and an age of 5$\pm$1 Gyr.
\cite*{pjm94} obtained not perfectly calibrated $BVI$ photometry and
measured a difference in magnitude between the main sequence turn-off point and
the red clump of $\delta V$=2.3, that translates, using the so-called MAI 
(Morphological Age Indicator, see \citealt{jp94}) into an age of 6.3 Gyr.
From their recalibration of the $\delta V$ - age relation, assuming 
[Fe/H]=$-0.23$, \cite*{salaris04} infer an age of 5.5 Gyr.
Note that the BDA\footnote{{\em http://www.univie.ac.at/webda//webda.html}} 
\citep{mermio95} indicates a spurious low age for this
cluster (1.1 Gyr), directly taken from the \cite{dias} catalogue, whose source
is unclear.
Finally, 
\cite{scott95} obtained low resolution spectra of 16 bright stars, from which 
an average cluster radial velocity (RV) was computed 
($\langle RV \rangle = -35 \pm 16$ km~s$^{-1}$). These spectra were later 
reanalyzed by \cite{friel02},  finding [Fe/H]=$-0.27$ (rms=0.15) dex. 

Be~32 has been photometrically studied by \cite{km91}, \cite{rs01} and 
\cite{pasj04}. Be~32 seems to be quite old (age about 6 Gyr) and moderately 
metal poor ([Fe/H] between -0.2 and -0.5).   
We have recently presented the RVs  of about 50 stars 
in Be~32 and a preliminary  analysis of the photometric
data \cite[hereafter D06]{dorazi06} based on isochrone fitting and
the magnitude of the red clump. 
In D06 we also discussed the
literature related to Be~32 available at the time, and we will not 
repeat it here. We now refine our determinations, applying the synthetic CMD 
method, as done for all the clusters in the BOCCE project. 
Finally, \cite{sestito} presented an analysis of high resolution FLAMES@VLT
spectra of 9 red clump giants in Be~32, finding an average metallicity
[Fe/H]$=-0.29$ dex (rms 0.04 dex), in very good agreement
with that found by D06.

\begin{table*}
\begin{center}
\caption{Log of observations for the clusters and the control fields; 
exposure times are in seconds.}
\begin{tabular}{lccrrrl}
\hline\hline
Field       & $\alpha_{2000}$ &$\delta_{2000}$   & $exp.time_B$ & $exp.time_V$ & $ exp.time_I$ & UT Date\\
\hline 
Berkeley 32   &$06^h58^m07^s$ &$+06^\circ 25'43''$  
 & 600,  40,  5 & 480,  20,  2 & 480,  20,  1 & 26/11/2000, 14/02/2004\\
Be 32 - ext   &$06^h57^m27^s$ &$+06^\circ 08'26''$  
 & 600, 240, 40 & 300, 120, 20 & 300, 120, 20 & 26/11/2000 \\
King 11       &$23^h47^m39^s$ &$+68^\circ 38' 25''$ 
 & 300, 1200, 240, 40 & 120, 600, 120, 20 & 120, 600, 120, 20& 25/11/2000, 26/11/2000\\
King 11 - ext &$23^h47^m40^s$ &$+68^\circ 08' 18''$ 
 &1200, 300, 40 & 600, 1280, 20 & & 25/11/2000\\ 
\hline
\end{tabular}
\end{center}
\label{tab-oss}
\end{table*}

The paper is organized as follows: observations and reductions are presented in
Sect. 2, a description of the  resulting CMDs can be found in Sect. 3; the
derivation of the cluster parameters using the synthetic CMD technique is
discussed in Sect. 4, while conclusions and summary are given in Sect. 5. 

\begin{figure}
\centering
\includegraphics[bb=180 50 420 740, clip, scale=0.75]{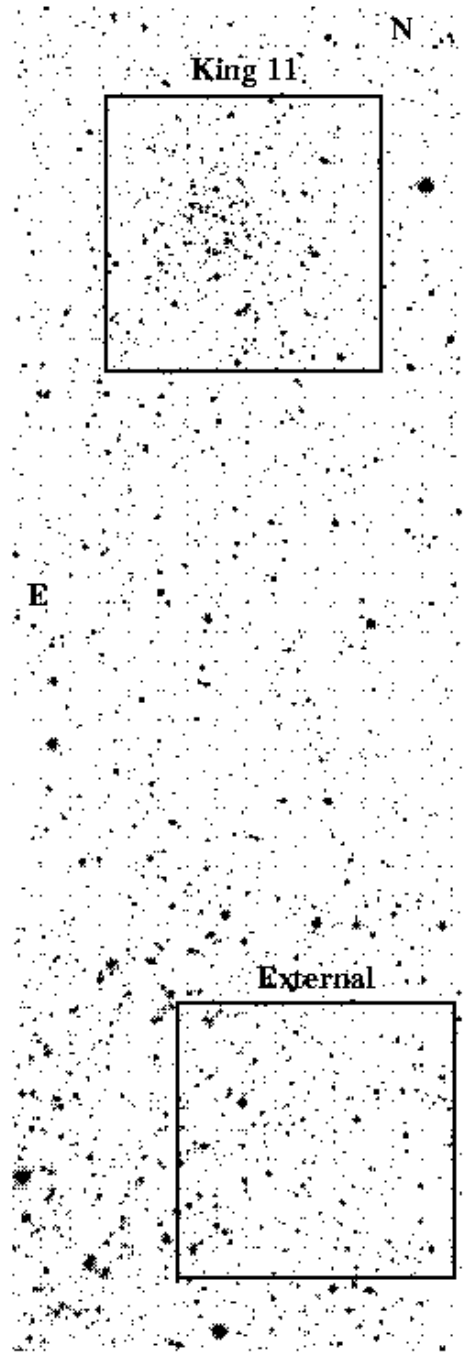} 	
\caption{
Approximate positions of our pointings on King~11 and the control field. 
The map is 15 $\times$ 45 arcmin$^2$, has North to
the top and East to the left.	
}	
\label{fig-map}
\end{figure}

\section{Observations and data reduction}

Observations in the $BVI$ Johnson-Cousins filters of Be\,32 and 
King\,11 were performed at the Telescopio Nazionale 
Galileo (TNG) in November 2000 (plus three additional exposures in 
February 2004 for Be~32). We also acquired  associated control fields to
check the field stars contamination,  as
detailed  in Table 1 and \cite{dorazi06}.  We used DOLORES (Device Optimized
for the LOw RESolution), with scale of 0.275 arcsec/pix,  and a field of view
9.4 $\times$ 9.4 arcmin$^2$.  Of the two November nights, only the first one
resulted photometric. Fig.~\ref{fig-map} shows the position of our pointings for
King~11 and the associated control field.

A description of the data and reduction procedure for Be~32 can
be found in \cite{dorazi05} and in D06; we report here briefly the analysis of 
King~11, which is absolutely equivalent to that of Be~32. The standard IRAF
\footnote{IRAF is distributed by the National Optical Astronomical
Observatory, which are operated by the Association of Universities for
Research in Astronomy, under contract with the National Science
Foundation } routines were utilized for pre-reduction, and the IRAF version of
the DAOPHOT-{\sc ii} package (\citealt{stetson87}, \citealt{davis94}) was used
with a quadratically varying point spread function (PSF) to derive positions
and magnitudes for the stars. Output catalogues for each frame were aligned in
position and magnitude, and final (instrumental) magnitudes were computed as
weighted averages of the individual values.
Even with the shortest exposure times we did not avoid saturation of the
brightest red giants in the $I$ filter; unfortunately, we could not obtain
additional exposures as we did for Be~32 (D06), so we will mostly concentrate
in the following on the $V,B-V$ CMD. 

The final catalogs have been created including all the objects identified in
at least two filters, after applying a moderate selection in the shape-defining
parameter $sharpness$  ($|sharpness| \leq 2$) and on the goodness-of-fit
estimator $\chi^2$   ($\chi^2 \leq 10$). To the two final catalogs, one for the 
cluster and one for the comparison field, we applied the transformation to
astrometrize the $\alpha$ and $\delta$ coordinates, using software written by P. Montegriffo at the
Bologna Observatory.

\begin{figure*}
\centering
\includegraphics[scale=0.8]{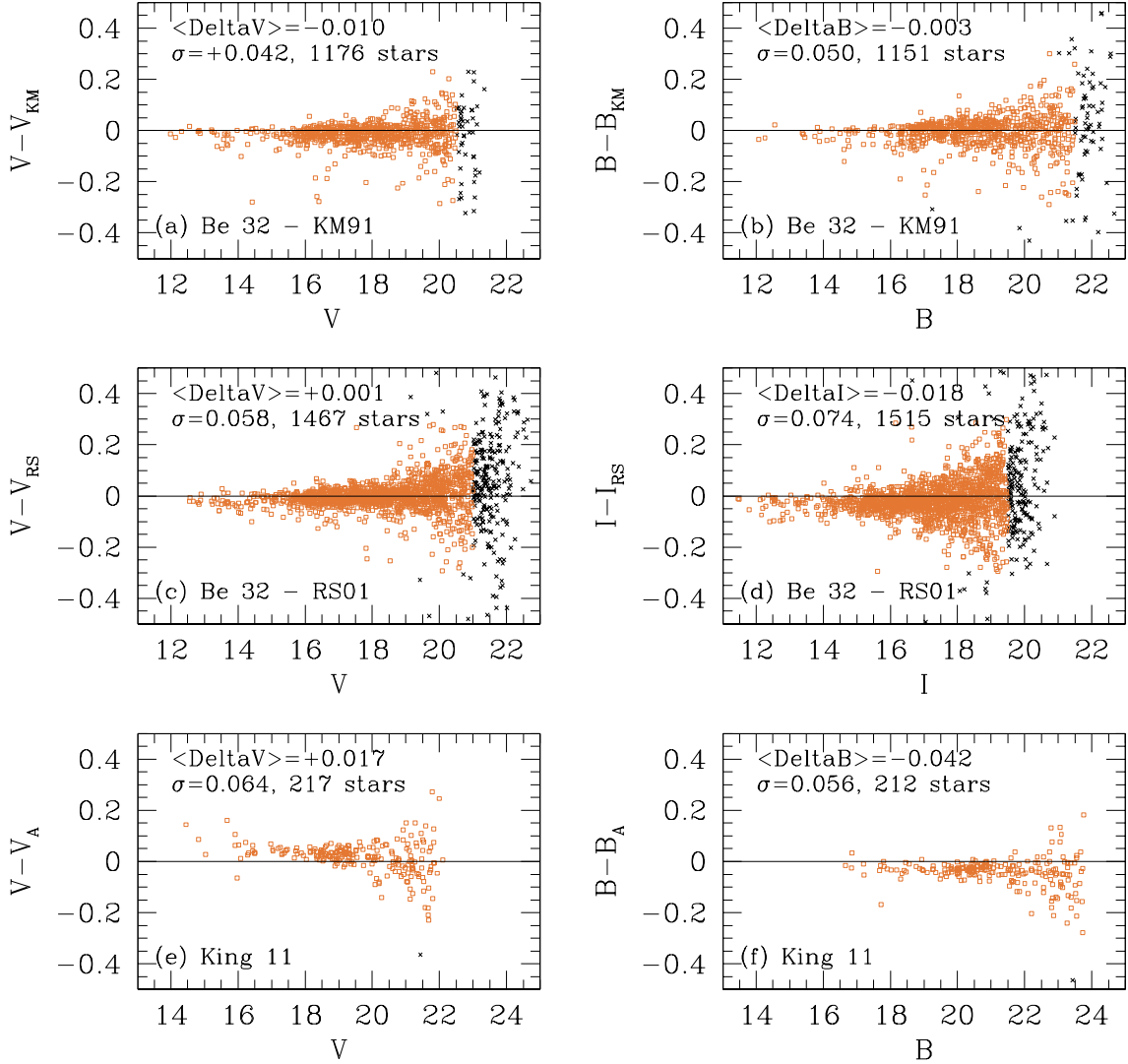} 
\caption{Comparison between our photometry and literature data.
(a) and (b) are for Be~32 by Kaluzny \& Mazur (1991); (c) and (d) are for Be~32
by Richtler \& Sagar (2001); (e) and (f) for King~11 by Aparicio et al. (1991).
The horizontal lines are on zero; stars used to compute the average
differences are indicated by (orange) open squares, while the ones discarded
are indicated by crosses.} 
\label{fig-conf}
\end{figure*}

After application of a correction to the PSF magnitudes to bring them on the
same scale of the aperture magnitudes of the standard stars, we calibrated
our catalogues to the standard Johnson-Cousins $BVI$ system.
We adopted the calibration equations that can be found in \cite{dorazi06}, 
since King~11 was observed in the photometric night beginning on
UT 2000 November 25 when Be~32 was observed too.

Finally, we determined our completeness level  using extensive artificial 
stars experiments: we iteratively added, one at a time, about 50000 simulated 
stars to the deepest frames and repeated the reduction procedure, determining the
ratio of recovered over added stars (see \citealt{tosi04} for a more detailed
description). The results for Be~32 are given in Table 2 
and those for  King~11  in Table 3.

We checked the quality of the calibration comparing our photometry for both
clusters with that presented
in previous literature papers, i.e. with \cite{km91}
for $B,V$ and with \cite{rs01} for $VI$ in Be~32, and with \cite{aparicio91}
for King~11 (only for $B,V$, since there are no other sources to compare 
the $I$ photometry with). 
Fig.~\ref{fig-conf} shows the differences with these
photometries for the stars in common; the comparison is particularly
favorable with the work by \cite {km91}, but is good in all cases.

\begin{table}
\begin{center}
\caption{Completeness level for the central (Cols 2, 3 and 4) and external
(Cols 5, 6 and 7) fields of Be~32; mag is the calibrated $B, V$ or $I$
  magnitude.}
\begin{tabular}{cllllll}
\hline\hline
mag & $c_B$ & $c_V$ & $c_I$ & $c_B$ & $c_V$ & $c_I$ \\
\hline
  16.00 & 1.00 & 1.00 & 1.00  & 1.00 & 1.00 & 1.00 \\
  16.50 & 1.00 & 0.95 & 0.92  & 1.00 & 0.99 & 0.95 \\
  17.00 & 0.92 & 0.94 & 0.88  & 0.99 & 0.98 & 0.94 \\
  17.50 & 0.91 & 0.93 & 0.85  & 0.97 & 0.97 & 0.92 \\
  18.00 & 0.89 & 0.92 & 0.78  & 0.97 & 0.94 & 0.87 \\
  18.50 & 0.88 & 0.91 & 0.68  & 0.96 & 0.93 & 0.84 \\
  19.00 & 0.86 & 0.87 & 0.54  & 0.93 & 0.93 & 0.73 \\
  19.50 & 0.82 & 0.85 & 0.37  & 0.91 & 0.90 & 0.52 \\
  20.00 & 0.77 & 0.80 & 0.21  & 0.89 & 0.86 & 0.29 \\
  20.50 & 0.66 & 0.74 & 0.09  & 0.85 & 0.78 & 0.11 \\
  21.00 & 0.51 & 0.60 & 0.03  & 0.69 & 0.58 & 0.04 \\
  21.50 & 0.32 & 0.39 & 0.01  & 0.42 & 0.32 & 0.01 \\
  22.00 & 0.16 & 0.19 & 0.00  & 0.22 & 0.15 & 0.00 \\
  22.50 & 0.06 & 0.09 & 0.00  & 0.07 & 0.05 & 0.00 \\
\hline
\hline
\end{tabular}
\label{complbe32}
\end{center}
\end{table}

\begin{table}
\begin{center}
\caption{Completeness level for the central (Cols 2 and 3) and external (Cols
  4 and 5) fields of King~11; mag is the $B$ or $V$ calibrated magnitude.  
}
\begin{tabular}{cllcll}
\hline\hline
   mag  & c$_B$  &c$_V$      & &c$_B$	&c$_V$    \\
\hline
 16.5  & 1.0 	&  1.0      & & 1.0    &  1.0    \\
 17.0  & 1.0 	&  0.99     & & 1.0    &  0.99   \\
 17.5  & 1.0 	&  0.97     & & 0.99   &  0.98   \\
 18.0  & 1.00	&  0.97     & & 0.98   &  0.95   \\
 18.5  & 1.00	&  0.95     & & 0.99   &  0.94   \\
 19.0  & 0.98	&  0.94     & & 0.96   &  0.94   \\
 19.5  & 0.97	&  0.93     & & 0.94   &  0.93   \\
 20.0  & 0.97	&  0.92     & & 0.91   &  0.90   \\
 20.5  & 0.97	&  0.87     & & 0.88   &  0.87   \\
 21.0  & 0.95	&  0.87     & & 0.81   &  0.82   \\
 21.5  & 0.93	&  0.74     & & 0.78   &  0.70   \\
 22.0  & 0.91	&  0.56     & & 0.63   &  0.43   \\
 22.5  & 0.88	&  0.27     & & 0.38   &  0.21   \\
 23.0  & 0.74	&  0.06     & & 0.15   &  0.04   \\
 23.5  & 0.45	&  0.00     & & 0.02   &  0.00   \\
 24.0  & 0.18	&  0.0      & & 0.00   &  0.0    \\
 24.5  & 0.02	&  0.0      & & 0.0    &  0.0    \\
 25.0  & 0.00	&  0.0      & & 0.0    &  0.0    \\
\hline
\end{tabular}
\label{king}
\end{center}
\end{table}

\begin{figure*}
\begin{center}
\includegraphics[bb=0 150 575 680, clip,scale=0.6]{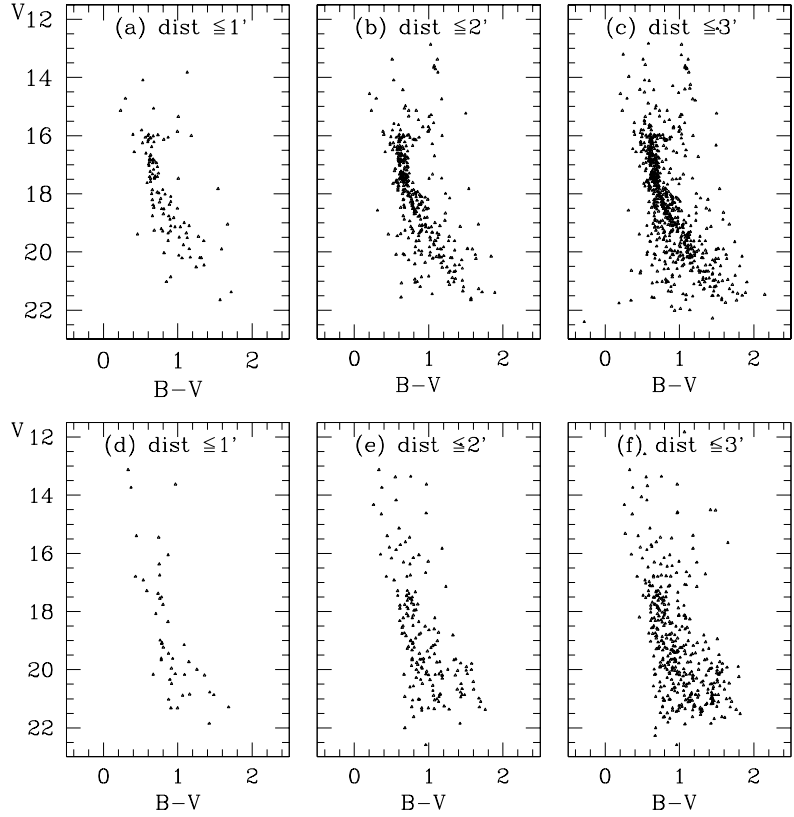} 
\caption{
Radial CMDs for Be~32 (upper panels) and equal areas in the
comparison field (lower panels); we plot stars within distances of 1, 2, 3 
arcmin from the cluster and field centres.
The CMDs contain 133, 444, 903 objects in panels (a), (b), (c) respectively,
and 57, 229, 524 in panels (d), (e), (f) respectively.}
\label{fig_rad_be}
\end{center}
\end{figure*}

\section{The colour - magnitude diagrams}

The CMDs for Be~32 were described in D06 
and the data are already available at the BDA. Fig.~\ref{fig_rad_be} 
shows the $V,B-V$ CMD of the stars at various distances from the centre of Be~32
and of the control field. It is apparent that contamination is quite high, with about
half the stars likely to be foreground/background objects even in the central
regions. However, in the area with a radius of 3$\arcmin$ from the 
cluster centre the main-sequence (MS), the turn-off (TO) and the subgiant 
branch (SGB) are  well defined. The MS extends more than 5 magnitudes below the
TO. With the additional help of the available RVs (from D06 and 
Randich et al. in preparation, see next section) to select the most probable 
cluster members, we can satisfactorily identify the TO 
($V=16.3$, $B-V=0.52$ and $V-I=0.60$), the SGB, the red giant branch (RGB), 
and the red clump ($V=13.7$, $B-V=1.07$ and $V-I=1.10$).  

\begin{figure*}
\centering
\includegraphics[bb=30 175 575 520, clip,width=15 cm, height=7 cm]{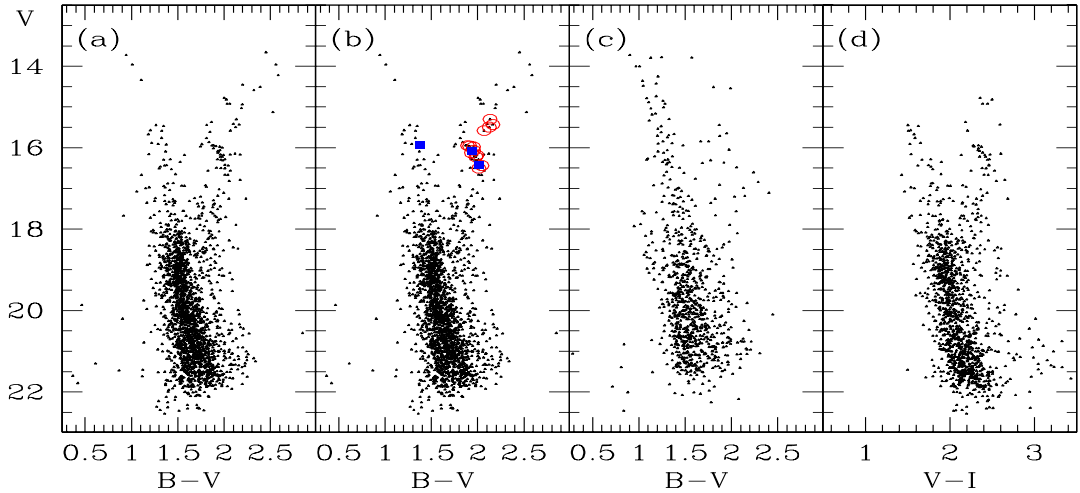} 
\caption{(a) $V, B-V$ CMD for King~11; (b) the same CMD, with stars member
(open circles, red in the electronic version) and non member (filled squares,
blue in the electronic version) according to the RVs in
Scott et al. (1995); (c)  $V, B-V$ CMD for the comparison field; (d)$V, V-I$ 
CMD for King~11 }	
\label{fig-cmd}
\end{figure*}

\begin{figure*}
\begin{center}
\includegraphics[bb=0 150 575 680, clip,scale=0.6]{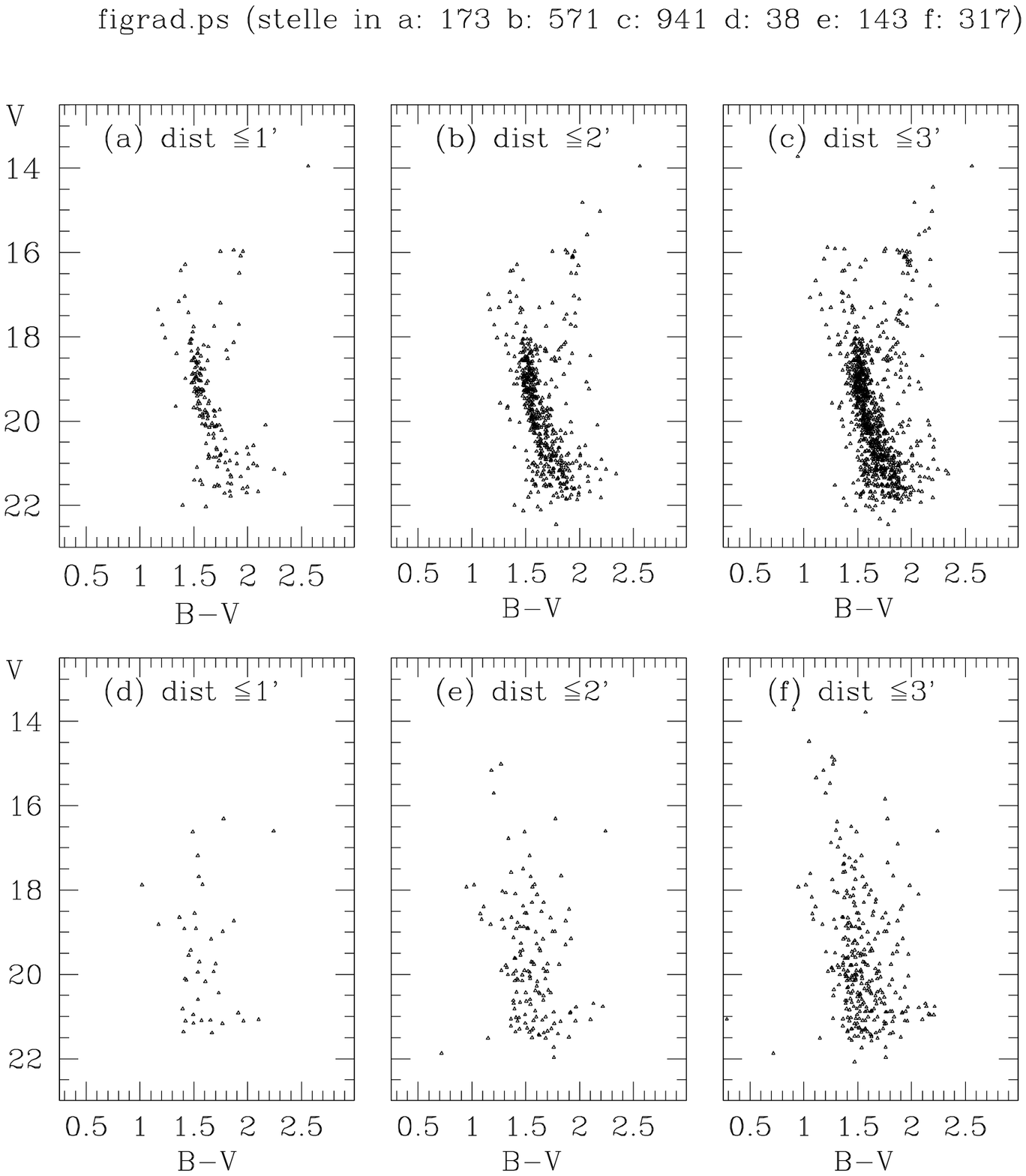} 
\caption{
Radial CMDs for King 11 (upper panels) and equal areas of the
comparison field (lower panels); we plot stars within distances of 1, 2, 3 
arcmin from the cluster and field centres.
The CMDs contains 173, 531, 941 objects in panels (a), (b), (c) respectively,
and 38, 143, 317 in panels (d), (e), (f) respectively.}
\label{fig-rad}
\end{center}
\end{figure*}

For King~11, the final, calibrated sample of  cluster stars (which will also be
made available through the BDA) consists of 1971 objects, and the external field
catalogue comprises 880 stars. The corresponding CMDs are shown in 
Fig.~\ref{fig-cmd}. In spite of a contamination lower than in Be~32, the location
of the foreground/background objects in the CMD makes the definition of the 
evolutionary sequences more complicated. We can improve the definition by using
the information on membership of a few giant stars from \cite{scott95}, which
perfectly define the red clump position. If we consider the CMDs of 
regions with increasing distance from the cluster centre displayed in 
Fig.~\ref{fig-rad}, it is apparent that a safe identification of the main
evolutionary loci becomes difficult beyond a radius of 2$\arcmin$. Within such
radius, the
cluster main sequence extends for almost 4 magnitudes and the RGB and red clump are well delineated. The Turn-off point is at
$V=18.2$, $B-V\simeq1.3$, while the red clump is at  $V=16.0$, 
$B-V\simeq1.8$.

In the $V,V-I$ CMD of King 11 we lack the brightest RGB stars, because 
they were saturated even in the shortest image, and the MS is less well defined.
For this reason, we refer to the $V,B-V$ CMD to derive the
cluster distance, reddening and age and 
use the $I$ data only to discriminate in metallicity among degenerate
solutions (see next Section).

\section{Cluster parameters}

Age, distance and reddening of King~11 and Be\,32 have been derived with the
same procedure applied to all the clusters of our project (see \citealt{bt06}
and references therein), namely the synthetic CMD method originally described
by \cite{tosi91}. The best values of the parameters are found by selecting the
cases providing synthetic CMDs with morphology, colours, number of stars in
the various evolutionary phases and luminosity functions (LFs) in better
agreement with the observational ones. As for the other clusters of this
series, to estimate the effect on the results of different stellar evolution
assumptions, we have adopted three different sets of stellar models, with
various assumptions on the metallicity, treatment of convection, opacities and
equation of state.  The adopted models are listed in Table 4.

\begin{table}
\begin{center}
\caption{Stellar evolution models adopted for the synthetic CMDs. The FST
models actually adopted here are an updated version of the published ones
(Ventura, private communication). }
\vspace{5mm}
\begin{tabular}{cccl}
\hline\hline
   Set  &metallicity & overshooting & Reference \\
\hline
BBC & 0.008 & yes 		    &Fagotto et al. 1994 \\
BBC & 0.004 & yes 		    &Fagotto et al. 1994 \\
BBC & 0.02  & yes 		    &Bressan et al. 1993 \\
FRA & 0.006 & no  		    &Dominguez et al. 1999 \\
FRA & 0.01  & no  		    &Dominguez et al. 1999 \\
FRA & 0.02  & no  		    &Dominguez et al. 1999 \\
FST & 0.006 & $\eta$=0.00,0.02,0,03 &Ventura et al. 1998\\
FST & 0.01  & $\eta$=0.00,0.02,0,03 &Ventura et al. 1998\\
FST & 0.02  & $\eta$=0.00,0.02,0,03 &Ventura et al. 1998\\
\hline
\end{tabular}
\end{center}
\label{models}
\end{table}

In addition to the usual synthetic CMD method, the cluster parameters have
also been searched by means of statistical tests. The problem of comparing
colour-magnitude diagrams (and two dimensional histograms in general) is still
unsolved in astrophysics. However, several approaches have been explored. For
instance, in \cite{cigno06} the {\it entire} CMD is used: data and model CMDs
are binned and a function of residuals is minimized. In \cite{gal99}, the
number of stars in {\it a few regions} (representative of the most important
evolutionary phases) is controlled through a $\chi^2$ test.  The goal of those
papers was to recover a complex star formation history. Here, the nature of
the problem is in principle simpler (single stellar generation), thus we
follow a more classical approach: the luminosity and the colour distribution
of each model are independently compared with the data using a
Kolmogorov-Smirnov (KS) test (\citealt{pr95}). One of the advantages of using
also the colour distribution lies in the fact that the major drawback of using
the LF alone, i.e, the degeneracy among parameters (distance, reddening, age
and metallicity) can be mitigated.  
Moreover, the KS does not require to bin the data; therefore, 
arbitrary parametrizations of the CMD (typical of the $\chi^2$) can be avoided. 
In order
to reduce the Poisson noise, that is the dominant uncertainty in our
luminosity functions, the model CMDs are built with a large number of
stars. 
Only CMDs yielding a KS probability larger than 5\% both for the LF and
for the colour distribution are accepted.

Unavoidably, poorly populated CMD regions like the core helium burning region or
the RGB are often under-represented by a similar analysis (washed
out by Poisson noise). However, also in these cases, a good KS probability still
indicates that the most populous stellar phases (e.g., MS and TO)
are well matched. In other words, the adopted statistical procedure provides a
quick tool to \emph{exclude} those solutions for which the synthetic CMD does 
not reproduce the properties of MS and TO stars. Then, the remaining parameter 
space is explored with a traditional analysis: 
i) exploiting the difference in luminosity between the lower envelope of
the subgiants and the red clump;
ii) fitting the SGB;
iii) matching the RGB colour.

\subsection{King 11}
As already said in Sect. 3, for King 11 we have mainly used the $V,B-V$ CMD
because the $V,V-I$  lacks the brighter part of the RGB.  To minimize
contamination from field stars we have selected as reference field the region
within a radius of 2\arcmin\, from its centre. Since this region contains 531
stars, and the control field of the same area contains 143 stars, we assume
the cluster members to be 388. Incompleteness and photometric errors are those
inferred from the data and described in Section 2. In order to minimize the
Poisson noise of the models, all available field stars ($\sim 880$) are used:
hence the synthetic CMDs are built with 3259 synthetic stars (in order to
preserve the ratio cluster members/field stars). Only afterwards we randomly
extract from the whole sample of synthetic stars 388 objects, as attributed to
the cluster central region.

\begin{figure}
\centering
\includegraphics[bb=70 190 575 640,scale=0.35,angle=270]{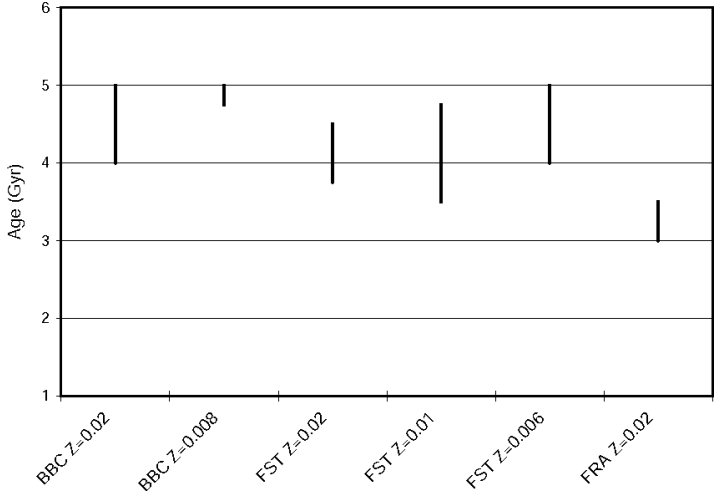}
\caption{The range of statistically acceptable ages for King 11. 
Results for different sets of tracks are shown.}
\label{exell}
\end{figure}

\begin{figure}
\centering
\includegraphics[scale=0.8]{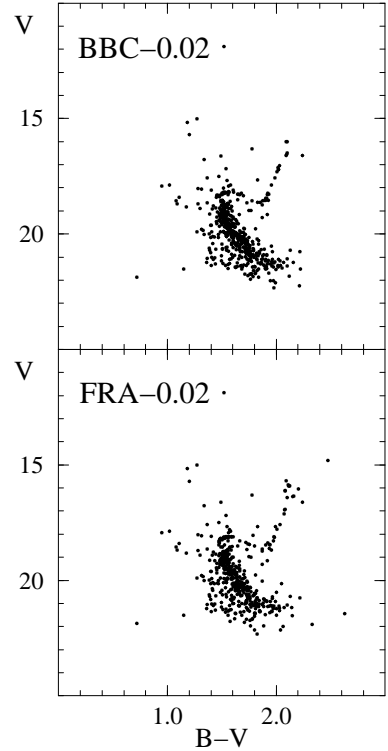} 
\caption{The upper panel shows the best solution for King 11 for BBC models: Z=0.02,
  $E(B-V)$=0.93 and (m-M)$_0$=11.85); 2) the lower panel shows the best FRANEC
  model: age=3 Gyr, Z=0.02, $E(B-V)$=1.01 and (m-M)$_0$=11.95). Both these models
  predict RGBs that are too red.}
\label{schifi}
\end{figure}

\begin{figure*}
\centering
\includegraphics[scale=.75]{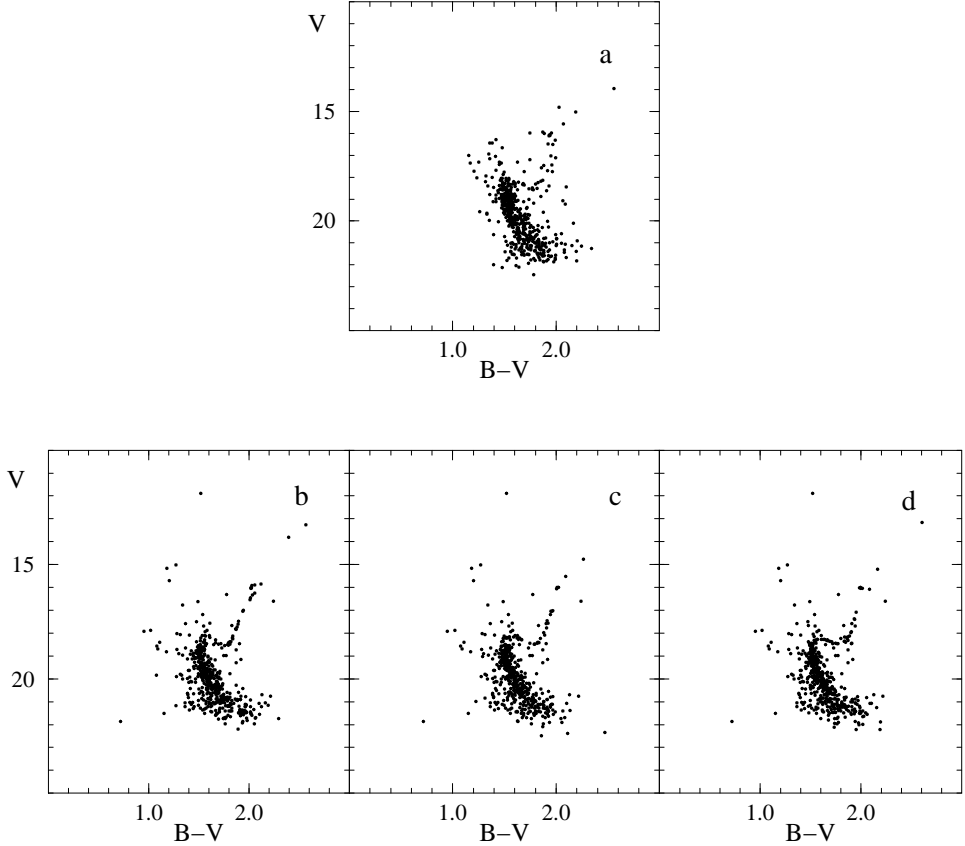} 
\caption{Comparison between observational and synthetic CMDs for King
  11. Panel a shows the data CMD for the central 2\arcmin ~radius
  region. Panels b, c and d show the CMDs of the best fitting cases (FST
  tracks): (b) age 4 Gyr, $E(B-V)$=0.94 and (m-M)$_0$=11.95, (c) Z=0.01, age
  4.25 Gyr, $E(B-V)$=1.04 and (m-M)$_0$=11.75, (d) Z=0.006, age 4.75 Gyr,
  $E(B-V)$=1.09 and (m-M)$_0$=11.65.}
\label{k11}
\end{figure*}

\begin{figure*}
\centering
\includegraphics[scale=.8]{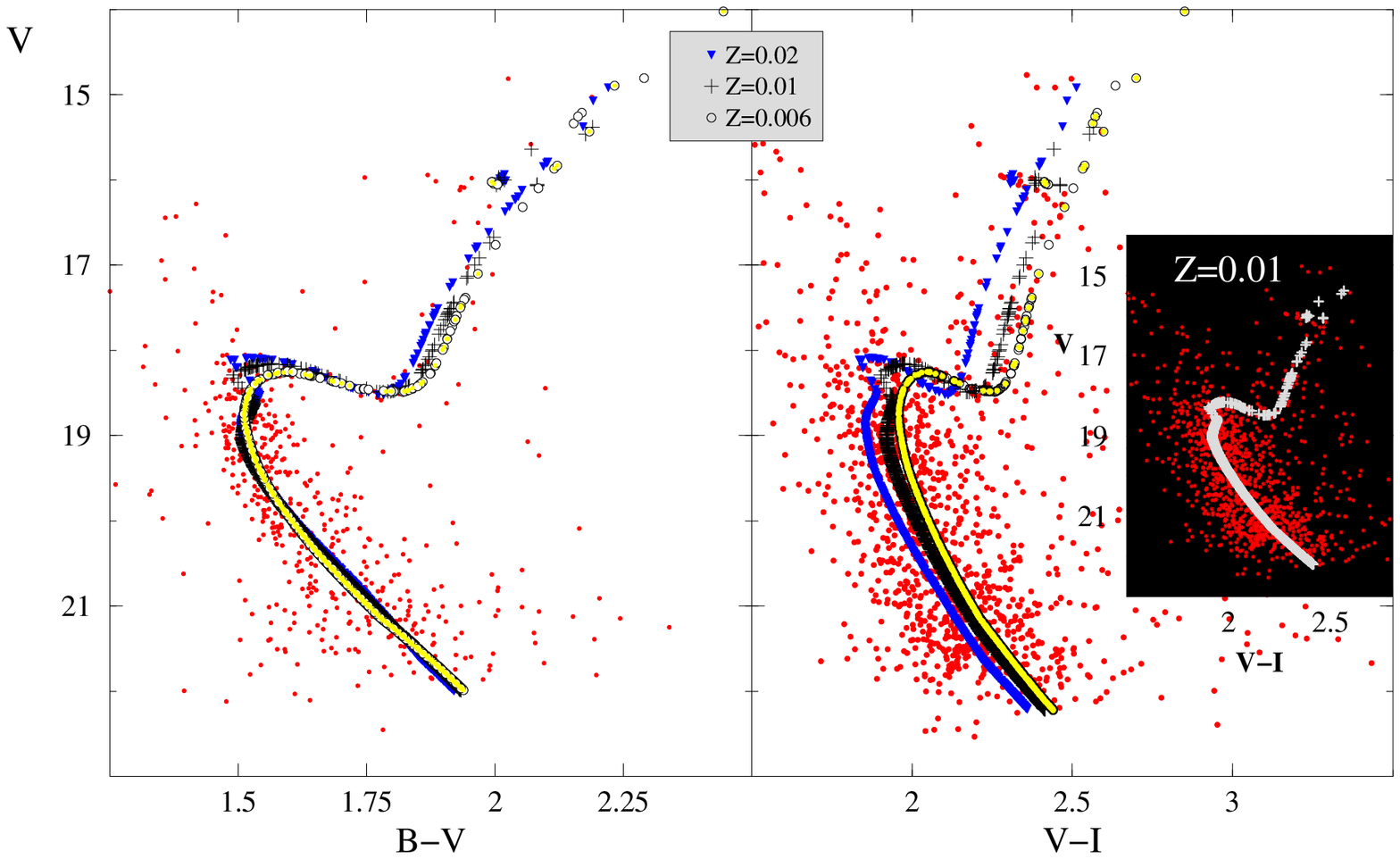} 
\caption{Choice of the metallicity for King~11: the left panel shows the
$V,B-V$ data and the three best solutions (at Z=0.006, 0.01, 0.02) that
all reproduce the observed CMD of the central zone, while
the right panel shows the same models overimposed on the $V,V-I$ data (in 
this case stars from the whole field are shown).
Only the solution at Z=0.01 (for an easier understanding it is isolated in the
small panel on the right) can well fit at the same time the two different
CMDs.}
\label{fig-bvi}
\end{figure*}

Almost all models have been computed assuming a fraction of binary stars of
20\% \footnote{The low number of observed TO stars doesn't permit to infer the
actual fraction.} (following Bragaglia \& Tosi 2006 prescriptions) and a power
law IMF with Salpeter's exponent.  The KS test is applied to the stars
brighter than $V\approx 20$.  The constraint on the KS probability doesn't
guarantee a unique solution, mostly because the statistics is dominated by MS
stars fainter than the TO, less affected than other evolutionary phases by
small parameters variations.  We have then decided to validate only models
with acceptable KS probabilities {\it and} with a predicted clump within
$0.05$ mag of the observed clump (whose membership is also confirmed by radial
velocity estimates).  Figure \ref{exell} shows the results \footnote{FRANEC
models for Z=0.006 and Z=0.01, providing the same age of Z=0.02, are not shown
in the figure.}; error bars correspond to ages for which an appropriate
combination of distance and reddening exists. Considering our findings, one
can provisionally accept a range of ages between 3 and 5 Gyr. Only BBC
models for Z=0.004 are rejected by the KS test for all ages (meaning that no
solution for age, reddening and distance has been found).
Figures \ref{schifi}, \ref{k11}, \ref{fig-bvi} show a selection of our best
synthetic CMDs. To further proceed in the selection, we have used the
morphology of the RGB (a poorly populated region, therefore ignored by our
statistical test) to give additional constraints on the parameter space. An
examination of this evolutionary phase reveals that: 1) the residual BBC
models (Z=0.02 and Z=0.008) are all rejected, because they predict excessively
red RGBs (the upper panel of Figure \ref{schifi} shows the best BBC model:
age=4.5 Gyr, Z=0.02, $E(B-V)$=0.93 and (m-M)$_0$=11.85); 2) the same problem
exists with the FRA models: the RGB is systematically too red (the lower panel
of Figure \ref{schifi} shows the best FRA model: age=3 Gyr, Z=0.02,
$E(B-V)$=1.01 and (m-M)$_0$=11.95); 3) the FST models seem in good agreement
with the data independently of the adopted metallicity.  We thus restrict the
next considerations only to the FST models.

Figure \ref{k11} shows the theoretical FST CMDs that best reproduce the $V,B-V$
data. The best fit parameters turn out to be: 
Z=0.02, age 4 Gyr, $E(B-V)$=0.94 and (m-M)$_0$=11.95 (panel a); 
Z=0.01, age 4.25 Gyr, $E(B-V)$=1.04 and (m-M)$_0$=11.75 (panel b); 
Z=0.006, age 4.75 Gyr, $E(B-V)$=1.09 and (m-M)$_0$=11.65 (panel c). 

To solve the degeneracy we have made use of the $V,V-I$ CMD: although not
complete in the bright part, it remains useful, since only models of the right
metallicity can fit the observed CMDs in all passbands (see also the case of
Be~32). Because of the very large reddening, we adopt the reddening law by
\citet[][see Appendix, eq. A1]{dean78}: 
$E(V-I)=1.25 \times E(B-V) \times[1+0.06(B-V)_0 +0.014 E(B-V)]$, 
which takes into account a colour dependence.
This relation tends to the usual $E(V-I) = 1.25\times E(B-V)$ 
 for $B-V \rightarrow 0$ and $E(B-V) \rightarrow 0$).

In Fig.~\ref{fig-bvi} we show the synthetic cases of Fig.~\ref{k11} both in
the $V,B-V$ and $V,V-I$ diagrams and with no photometric error, to allow for a
more immediate visualization of the theoretical predictions. We can see from
Fig.~\ref{fig-bvi} that the three competing models, indistinguishable in $B-V$
(left panel), do separate in $V-I$ (right panel): the best fit is reached for
Z=0.01. The solar composition seems definitely ruled out (the MS is too blue),
but the Z=0.006 model lies only slightly too red and cannot be completely
excluded.  This seems to confirm the findings by \cite{friel02}, who based the
analysis on spectroscopic indices. In contrast, \cite{aparicio91} preferred a
solar abundance on the basis of their CMDs, but in this case different stellar
models have been employed.  While we are rather confident on a subsolar
metallicity, a definitive answer will require analysis of high resolution
spectra.

The assumption  of different levels of core overshooting ($\eta$ = 0.2 or 0.3) 
has a minor effect
on the results, as expected: King 11 is a sufficiently old cluster that the
upper MS stars have masses with small convective cores, and therefore with small
overshooting. Comfortably, the predicted number of stars in RGB and clump phase
is close to the observed one, confirming that the evolutionary lifetimes of the
theoretical models are correct. 

Finally, in order to evaluate the contribution of the adopted binary fraction
and IMF, we performed several tests. Larger fractions of binaries could help to
fit the MS, yielding slightly larger distance moduli 
(with minor effects on
the age). Viceversa, if distance, reddening and age are fixed, the stellar
multiplicity that is consistent with the data is wide (between 10\% and
60\%). In fact, only fractions higher than 60\% produce an evident plume over
the turn-off region, which is not observed. If the same test (fixing distance,
reddening and age) is performed also for the IMF, the results allow to rule
out only exponents larger than 3.3, for which the synthetic RGBs appear
underpopulated.

In conclusion, the best parameters for King 11
can be summarized in the following intervals:
\begin{itemize}
\item Z=0.01;
\item age between 3.5 to 4.75 Gyr;
\item distance modulus between 11.67 and 11.75;
\item reddening $1.03 \le E(B-V) \le 1.06$.
\end{itemize}

\begin{figure*}
\centering
\includegraphics[scale=0.8]{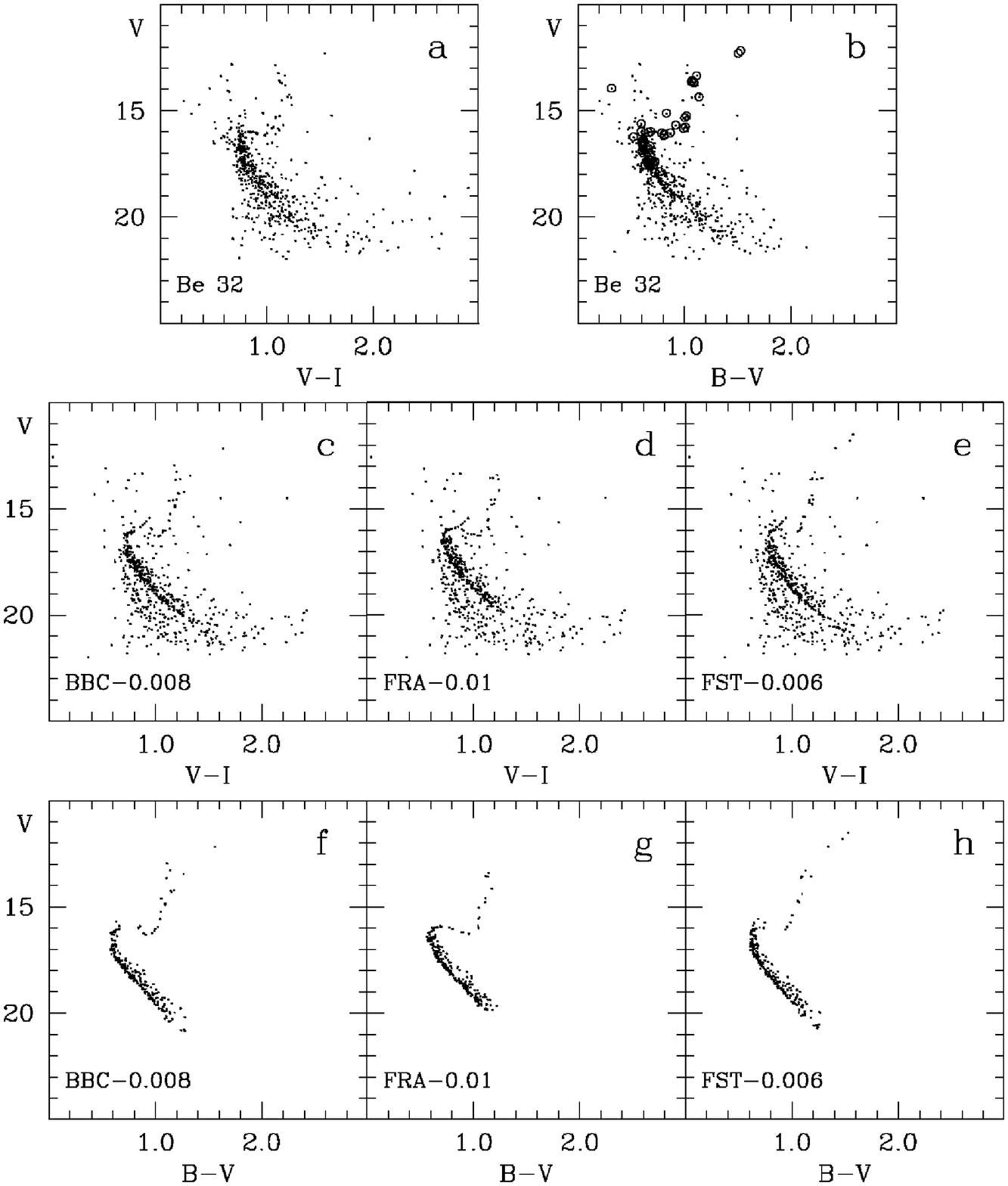} 
\caption{Comparison between observational and synthetic CMDs for Be~32. Panels
a and b show the stars measured in $B, V, I$ in the central 3\arcmin ~radius
region. The larger symbols (red in the electronic version) in panel b indicate the objects with higher
membership probability from the RVs (see text for details). Panels f, g and h
show the $B-V$ CMDs of the best fit case, mentioned in the text, for each
type of stellar models. Panels c, d and e show the corresponding $V-I$ CMDs,
overimposed to the CMD of the same area in the control field for a more direct
comparison.}
\label{besynt}
\end{figure*}

\subsection{Berkeley 32}
For Be~32, we have chosen as reference CMDs those of the region within 3\arcmin 
~from the cluster centre (top panels in Fig.~\ref{besynt}), which contains 608
stars with magnitudes measured in all the three $B, V, I$ bands. The same area
in the control field contains 332 stars with $B, V, I$. Taking this
contamination into account, as well as the circumstance that 27 of the stars
within the central area are shown by the RVs not to belong to Be~32, we
assume the cluster members to be 249. 
The top panel of Fig.~\ref{besynt} shows the CMD of
the stars located within 3\arcmin ~from the cluster centre, with the larger
symbols indicating the 48 objects whose RVs indicate most probable membership. 
To help
in the RGB definition, also the two brightest RGB members are shown, although
outside the selected 3\arcmin ~radius.

The synthetic CMDs have been generated with 249 objects, the incompleteness of
Table 2 and the photometric errors described by D06. We have
generated the synthetic CMDs with and without binary systems.  As for most of
our sample clusters, a fraction of 30\% of binaries seems more consistent
with the data, for all sets of stellar models. We notice, though, that
binaries are not sufficient to cover the whole colour extension of the MS: a
differential reddening of about $\Delta E(B-V)=\pm0.01$ would provide a better
reproduction of the MS thickness.

The results of our
analysis are the following. A solar metallicity is out of the question,
because the synthetic CMDs show $V-I$ colours definitely too blue for all
cases when the $B-V$ colours are correct. Of all the synthetic models, only
those with metallicity Z=0.008 are always able to simultaneously reproduce 
both the $B-V$ and $V-I$ colours of all the evolutionary phases. For Z$<$0.008, 
if $B-V$ is reproduced, $V-I$ tends to be too red, while for Z$>$0.008, if 
$B-V$ is fine, $V-I$ tends to be too blue. Unfortunately, Z=0.008 is available 
only for the BBC tracks. For the FRA models, an acceptable colour agreement is
achieved for Z=0.006, but when we take into account also the shape of the MS
and the TO, Z=0.01 may be better. With the FST models, instead,  Z=0.006
seem slightly better than Z=0.01. This ambiguity further suggests
that the actual metallicity is in between, i.e, Z=0.008.

In order to obtain an in depth exploration of the preferred metallicity
Z=0.008, we have also applied our statistical procedure. Although the
contamination by field stars is quite high, the turn-off region, also thanks
to the partial cleaning from non members by the RVs, appears better defined
than in King 11. The KS test is simultaneously applied to the V, B-V and V-I
distributions, selecting only models giving a KS probability above 5
percent. The only acceptable models resulted to have age between 5 and
6.1 Gyr, distance moduli $(m-M)_0=12.5-12.6$ and reddening
$0.085<E(B-V)<0.12$. 

Whatever the metallicity, it is not easy to reproduce the shape of all
the evolutionary phases covered by the stars in Be~32. The BBC models, in spite
of the excellent reproduction of the colours, shape and position of MS, SGB and
RGB, do not fit precisely the morphology of the TO and predict a 
clump slightly too
bright. The FRA models are the only ones with a TO hooked enough to fit 
the bluest supposed member of Fig.~\ref{besynt} (which however is
in the tail of the RV distribution and is the least safe member), but not for
the ages which better reproduce the other CMD sequences. When the TO morphology
is fine, the clump is too bright and vice versa. Moreover, the MS of the FRA
models is slightly too red at its faint end. 
The FST models, independently of the overshooting choice $\eta$,
have TO not much  hooked  and 
excessively vertical RGBs, whose brightest portion is therefore too blue. 

As usual, models without overshooting (FRA) lead to the youngest age.
The FST models with maximum overshooting $\eta$=0.03 provide results totally
equivalent to those with $\eta$=0.02; this has been noted also for King~11 and
all OC's old enough to have stars with small (or no) convective cores.  

The best compromise for each set of stellar models is: 
\begin{itemize}
\item Z=0.008, age 5.2 Gyr, $E(B-V)$=0.12, (m-M)$_0$=12.6 (BBC); 
\item Z=0.01, age 4.3 Gyr, $E(B-V)$=0.14, (m-M)$_0$=12.6 (FRA); 
\item Z=0.006, age 5.2 Gyr, $E(B-V)$=0.18, (m-M)$_0$=12.4 (FST). 
\end{itemize}
 The CMDs
corresponding to these three best cases are shown in Fig.~\ref{besynt}, where
in $V, B-V$ we plot only the synthetic stars to allow for a direct comparison
of the different models, while in $V, V-I$ we overplot the control field
objects on the synthetic stars to facilitate the comparison between
theoretical and observational CMDs.

The uncertainties mentioned above obviously affect the identification 
of the best age; however, all our independent tests consistently favour an 
age between 5.0 and 5.5 Gyr with
overshooting models (both BBC and FST, although the BBC ones perform better,
possibly because of the more appropriate metallicity Z=0.008).

Finally, another useful piece of information can be inferred from the
comparison of the pure synthetic CMDs of the bottom panels of
Fig.~\ref{besynt} with the observational ones of the top panels. 
The  synthetic MSs don't reach magnitudes fainter than $V\simeq$21 for BBC and
FST and  $V\simeq$20 for FRA. This limit corresponds to the minimum
stellar mass available in the adopted sets of models: 0.6\msun\, in the BBC
and FST sets and 0.7\msun\, in the FRA ones.  In the central row panels, 
where the
external field CMD is overimposed to the synthetic one, the faintest portions
are therefore populated only by foreground/background stars. Yet, the
synthetic LFs don't differ too much from the observational one, suggesting
that contamination dominates at that magnitude level.

\begin{table*}
\begin{center}
\caption{Comparison of our results and selected literature data for
the two clusters. }
\begin{tabular}{lccccl}
\hline\hline
Authors       & age (Gyr)& Z or [Fe/H] & $(m-M)_0$& E(B-V)& Notes\\
\hline
&&&{ \large King 11}&&\\
\hline
This work        & 3.5-4.75 &0.01           & 11.67--11.75& 1.03-1.06      &$BVI$ \\
Kaluzny          &$\sim5$   &               &    $(m-M)_V\sim15.3$        &  &Shallow $BVR$, comparison to M67/red clump mag\\
Aparicio et al.  &$5\pm1$   &0.02           & 11.7    &   1.00               & $BVR$, synthetic $V,B-V$ CMD  \\
Salaris et al.   & 5.5      &$-0.23\pm0.15$ &           &                    & $\delta V$, [Fe/H] from liter., age-metallicity-$\delta V$ relation  \\
\hline
&&&{ \large Berkeley 32}&&\\
\hline
This work        & 5.0-5.5 & 0.008          &12.4--12.6       & 0.12   & $BVI$\\
Kaluzny \& Mazur &6      & $-0.37\pm0.05$ & 12.45$\pm$0.15 & 0.16      & Morphological Age Ratio/MS fitting \\
D'Orazi et al.   & 6.3  & 0.008           &12.5--12.6	   & 0.10      & $BVI$, isochrone fitting/red clump mag\\
Richtler \& Sagar&6.3    & $-0.2$         & 12.6$\pm$0.15  & 0.08      & $VI$, isochrone fitting/red clump mag\\
Sestito et al.   &       & $-0.29\pm0.04$ &                & 0.14      & High-res spectra \\
\hline
\end{tabular}
\end{center}
\label{sum}
\end{table*}

\section{Summary and discussion}
The context of this work is the large BOCCE project (\citealt{bt06}), devoted
to the systematic study of the Galactic disc through open clusters. Distance,
reddening and physical properties of the open clusters King 11 and Be 32 have
been explored. To this end, synthetic CMDs have been built and compared with
data using both morphological and statistical criteria. A morphological analysis exploits
all the evolutionary phases, but leads to some level of subjectiveness. On the other hand, a pure statistical treatment can establish
the significance for each model (reducing the subjectiveness of the
comparison), but is truly selective only in case of very well defined TOs.

In order to extract the maximum level of information, we have used both
 approaches: 1) we generate synthetic CMDs to best reproduce the
 main CMD features, especially the late evolutionary phases (RGB, red clump
 luminosity, SGB); 2) TO and main sequence are explored by KS test (LF and
 colour distribution). The final results come from the intersection of these.

During the analysis, King 11 and Be 32 have presented different problems. For
King 11, whose metallicity is unknown, the statistical treatment has the
advantage to explore very quickly a multidimensional parameter
space. Nevertheless, King 11 has a very noisy TO, therefore, a morphological
analysis plays a key role in refining the results. On the other hand, Be~32 is
characterized by well defined TO and MS (and a well defined metallicity), and
the statistical approach has provided an independent estimate of the parameters.

For King 11, our analysis has produced the following results: (1) the FST
tracks give the best chance to reproduce the LF, the colour distribution and
the morphological constraints (the clump luminosity, the bottom of the RGB and
the RGB colour); (2) the metallicities Z=0.006, Z=0.01, Z=0.02 all
produce synthetic $V,B-V$ CMDs whose goodness of fit 
are indistinguishable but the use of the $I$ band permits to select the right cluster metallicity,
i.e. Z=0.01; (4) the synthetic CMDs generated with the FST tracks are consistent
with a reddening $1.03 \le E(B-V) \le 1.06$, a distance modulus between
11.67 and 11.75, a cluster age between 3.5 and 4.75 Gyr (the best fit is
obtained with 1.04, 11.75 and 4.25, respectively).

Our results confirm that King 11 is among the true ``old open cluster'',
contradicting the \cite{dias} value, but in line with all past
direct determinations. For an immediate comparison,
Table 5 shows our results together with  literature ones.  Our derived
ages are consistent with the \cite{aparicio91} finding (age $5 \pm 1$
Gyr). The difference (our estimates are systematically younger) may be
easily ascribed to the input physics: \cite{aparicio91} adopted the
\cite{bressan81} tracks, characterized by strong core overshooting: although
King 11 masses are only marginally affected by this phenomenon, a conspicuous
amount of overshooting goes in the direction of rising the estimated age. A
similar age is recovered also by \cite{kaluzny89}, but that work is
based on a very shallow sample.  \cite{salaris04}, adopting [Fe/H]=$-0.23$,
provide an age of about 5.5 Gyr from their recalibration of the relation
between $\delta V$, metallicity and age, based on ten clusters. The
large reddening we have found is in good agreement with literature values, in
particular with the $E(B-V)=0.98$ derived by the \cite{sfd98} maps. Our choice
of metallicity is in good agreement with the one by \cite{friel02} and
slightly discrepant with the other derivation based on photometry
\citep{aparicio91}, which, however, is more uncertain since those authors
found discrepant results with different methods.

In the case of Be~32 our CMDs constrain fairly well the cluster metallicity. The
BBC tracks for Z=0.008 reproduce all the stellar phases in all bands, while
other metallicities have problems to simultaneously best fit
both the $V, B-V$ and the $V, V-I$ diagrams.  This is in perfect agreement
with the finding by \cite{sestito}, based on high resolution spectra
([Fe/H]$=-0.29 \pm 0.04$).  The best estimate of the age ranges between 5.0
and 5.5 Gyr, slightly older than King~11. The age derived by D06 with isochrone fitting
was 6.3 Gyr, consistent with what we find here once we consider
the coarseness of the isochrone grid. Slightly older ages (6.3 and 6.0 Gyr,
respectively) were found also by \cite{rs01} and \cite{km91}, while 
\cite{pasj04} reach exactly our same conclusion (5.2 Gyr).

In addition, the present data for Be 32 suggests a distance modulus 
$(m-M)_0=12.4-12.6$, in fair agreement with past
studies,  and reddening most likely around 0.12.
The latter is consistent but slightly larger than the $E(B-V)=0.10$ we 
determined in D06 assuming an older age, and slightly smaller than the value $E(B-V)=0.16$
quoted by \cite{km91}. A clearly lower reddening
($E(B-V)=0.08$) was found by \cite{rs01}, but we recall that their study
was based only on two passbands and may be plagued by uncertainties like
the ones we found in the case of our analysis of King~11.
The comparison to the \cite{sfd98} maps is too uncertain, given the very low
latitude of the cluster. 
We suggest the possibility of a differential reddening of the order of
$\Delta E(B-V)\simeq$0.02.

We have computed the distances of the two OCs adopting the preferred distance
moduli: King~11 has a distance of about 2.2-3.4 kpc from the Sun and
about 9.2-10 kpc from the Galactic centre (assuming the Sun to be at 8 kpc from the
centre), with a height above the Galactic
plane of 253-387 pc; the corresponding values for Be~32 are 3.0-3.3 kpc, 10.7-11 kpc, and
231-254 pc, respectively. Neither cluster is far enough from the Galactic centre
to  be of relevance in the current debate about the metallicity distribution
in the outer disc. However, both contribute to enlarge the still smallish
number of old OCs and their metallicity (specially once that of King~11 is
confirmed by dedicated high resolution spectroscopy studies) will be important
in defining the (possible) variation of the radial metallicity distribution over
the Galactic disc lifetime.

\bigskip\noindent ACKNOWLEDGEMENTS \par\noindent The King 11 data
reduction was performed by Roberto Gualandi of the Loiano Telescope staff.  We are grateful to
Sofia Randich for the RVs of Be~32 provided in advance of publication.  
We gratefully
acknowledge the use of software written by P. Montegriffo, and of the BDA
database, created by J.C. Mermilliod, and now operated at the Institute for
Astronomy of the University of Vienna. This project has received partial
financial support from the Italian MIUR under PRIN 2003029437.

\end{document}